# Title: Extraction of Isotropic Electron-Nuclear Hyperfine Coupling Constants of Paramagnetic Point Defects from Near-Zero Field Magnetoresistance Spectra via Least Squares Fitting to Models Developed from the Stochastic Quantum Liouville Equation

Authors: Elias B. Frantz[1], Nicholas J. Harmon[2], Stephen R. McMillan[3], Stephen J. Moxim[4], Michael E. Flatté[3], Patrick M. Lenahan[4,1],

Affiliation: [1]Intercollege Graduate Degree Program in Materials Science and Engineering, Pennsylvania State University, University Park, PA 16802, USA

[2]Department of Physics, University of Evansville, Evansville, IN, 47722, USA

[3]Department of Physics and Astronomy, University of Iowa, Iowa City, IA, 52242, USA

[4]Department of Engineering Science and Mechanics, Pennsylvania State University, University Park, PA 16802, USA

## Abstract

We report on a method by which we can systematically extract spectroscopic information such as isotropic electron-nuclear hyperfine coupling constants from near-zero field magnetoresistance (NZFMR) spectra. The method utilizes a least squares fitting of models developed from the stochastic quantum Liouville equation. We applied our fitting algorithm to two distinct material systems: $Si/SiO_2$ metal oxide field effect transistors (MOSFETs), and a-Si:H metal insulator semiconductor (MIS) capacitors. Our fitted results and hyperfine parameters are in reasonable agreement with existing knowledge of the defects present in the systems. Our work indicates that the NZFMR response and fitting of the NZFMR spectrum via models developed from the stochastic quantum Liouville equation could be a relatively simple yet powerful addition to the family of spin-based techniques used to explore the chemical and structural nature of point defects in semiconductor devices and insulators.

## I.) Introduction

For decades, electron paramagnetic resonance (EPR) has been used to determine the physical and chemical nature of paramagnetic point defects in semiconductor devices and insulators.[1–7] When studying fully formed micro- and nanoscale electronic devices, EPR is limited in both its sensitivity and selectivity to defects directly involved in device performance. A closely related technique, electrically detected magnetic resonance (EDMR), takes advantage of spin-dependent currents in these devices to overcome these shortcomings.[8,9] However, EDMR spectrometers are complex and costly. EDMR is also still limited in its potential to study devices below metallization layers in three-dimensional integrated circuits. A new technique, near-zero-field magnetoresistance (NZFMR), has recently been investigated as a new type of spectroscopy that is capable of overcoming these issues to study point defects in fully processed devices.[10]

The NZFMR measurement is simpler than both EPR and EDMR measurements. The measurement may provide much of the analytical power of the resonance measurement while also being a suitable technique to study point defects in fully processed three-dimensional integrated circuits.[10] In this work, we discuss a theoretical approach to the analysis of the NZFMR spectrum which can provide values for electron-hyperfine interactions in multiple systems. These hyperfine values are sufficiently accurate to provide substantial physical insight with some simplifying assumptions. The approach utilizes modeling of the stochastic Liouville equation (SLE). We have developed a nonlinear least squares fitting algorithm based on modeling of the SLE to extract spectroscopic information from NZFMR spectra. We applied the algorithm to two distinct material systems: $Si/SiO_2$ metal oxide field effect transistors (MOSFETs), and a-Si:H metal insulator semiconductor (MIS) capacitors.

In order to better understand the NZFMR measurement, a basic understanding of EPR and EDMR is useful. We first consider a free electron placed in a slowly varying magnetic field, $B_0$, and exposed to a perpendicular oscillating magnetic field, $B_1$, with frequency $\nu$. The slowly varying magnetic field provides an energy splitting between the spin-up and spin-down state of the electron, known as Zeeman splitting. The resonance condition of the electron is reached when the energy difference between the two spin states is equal to the product of Planck's constant by the frequency of the perpendicular oscillating magnetic field. The resonance condition for the simplest case is of an unpaired electron otherwise isolated from its surroundings is given by

$$h\nu = g_e \mu_B B_0, \tag{1}$$

where $h$ represents Planck's constant, $\nu$ is the frequency of the perpendicular oscillating magnetic field, $g_e$ is the Lande g-factor, $\mu_B$ is the Bohr magneton, and $B_0$ is the magnitude of the applied magnetic field.

The analytical power of EPR comes from deviations to this resonance response when the electron resides in paramagnetic defect sites, in real material systems. In such cases, the EPR response is altered by the local environment. In most cases, the two most important factors that alter the resonance condition are spin-orbit coupling and electron-nuclear hyperfine interactions. Spin-orbit coupling changes the isotropic Landé g-factor (2.0023…) to an orientation-dependent factor generally expressed as a second rank tensor. Electron-nuclear hyperfine interactions between the electron in the paramagnetic defect site and nearby nuclei with magnetic moments also alter the resonance response. These hyperfine interactions result in a splitting of the energy levels of the system which is dependent on both the spatial distribution of these magnetic nuclei, and the wavefunction of electrons in the defect under observation. Considering both spin-orbit coupling and electron-nuclear hyperfine interactions with nearby magnetic nuclei, the resonance condition becomes

$$h\nu = g\mu_B B_0 + \sum_i m_i A_i. \qquad (2)$$

The free electron $g_e$ is replaced by an orientation-dependent $g$ which is usually expressed as a second rank tensor, $m_i$ is the nuclear spin quantum number of the $i^{th}$ nearby magnetic nuclei, and $A_i$ is the electron-nuclear hyperfine coupling due to the nucleus usually expressed as a second rank tensor.

In the EPR measurement, a sample containing paramagnetic defects is placed in a high-Q microwave cavity which is tuned to the cavity's resonance frequency. The cavity sits in an electromagnet which slowly varies the magnetic field across the resonance condition. In classical EPR, at the resonance condition, an absorption of power is detected. This microwave absorption is plotted as a function of the magnetic field. Accurate measurement of both the frequency of the oscillating magnetic field and the magnitude of the slowly swept magnetic field allows for the evaluation of g-tensor and hyperfine tensors. The resonance condition is highly dependent upon the defect's local surroundings; thus, the identification of the chemical and physical nature of the defects is possible with such a measurement.

Conventional EPR measurements are sensitive to about $10^{10}$ total paramagnetic defects[11], greatly limiting its application to the study of defects in state-of-the-art micro- and nanoscale devices. EPR is not necessarily able to directly determine which defects are directly involved in device performance because the technique is sensitive to all paramagnetic centers within the sample. Additionally, the observations of such defects often requires measurements in fully processed devices with far fewer than $10^{10}$ total defects. EDMR is a variation of EPR which eliminates these shortcomings. EDMR accomplishes this through the measurement of EPR-induced changes in device current as a function of the magnetic field rather than a change in microwave absorption. The majority of EDMR studies have utilized one of two mechanisms: spin-dependent recombination (SDR)[8,12–20] and spin-dependent trap assisted-tunneling (SDTAT)[9,21–24]. Each of these methods will be discussed in detail in the modeling section of this paper.

In trying to understand defects present in three-dimensional integrated circuits, both classical EPR and EDMR are limited by their use of a microwave or radio frequency magnetic field which is perpendicular to the slowly varying magnetic field. The penetration depth of an oscillating magnetic field is a strong function of frequency. As frequencies are increased, penetration depth decreases. Performing EPR and EDMR measurements at lower frequencies could overcome this problem. However, in the construction of three-dimensional integrated circuits, metallic interlayers placed in between the layers of the devices of interest will completely shield these devices from any oscillating magnetic field. Without the perpendicular oscillating magnetic field, both techniques are impossible.[10]

During a standard EDMR measurement, a change in current can also be measured when the magnetic field is swept through zero: the NZFMR response.[8–10,23,25–29] The NZFMR response occurs with or without the presence of the oscillating magnetic field. The elimination of the oscillating magnetic field makes the measurement much simpler than conventional EPR and EDMR. The small magnetic fields utilized in NZFMR also simplify the measurement. In some systems, the NZFMR response has been shown to closely scale with the amplitude of the EDMR response to the EDMR response in studies of device stressing[25], radiation damage[10,23,29], changes in temperature[26] and bias.[9,10,26] Although some analytical results have been extracted from the NZFMR spectrum, no systematic approach has been available with which to obtain information on the strength of the interaction given a defined spin cluster. We report on a systematic method to extract information about electron-nuclear hyperfine interactions from the NZFMR via a least squares fitting method. We do this using solutions of the SLE. We show that it is possible to extract information about hyperfine parameters utilizing this approach in the two different material systems: MOSFETs and MIS capacitors.

## II.) Modeling NZFMR Responses

Several mechanisms have been explored to explain similar magnetic-field effects on current in organic semiconductors.[30–35] The proposed models for these similar effects in organics focus on the possible spin-spin interactions present in bipolaron transport[32,36], electron-hole (e-h) pair recombination[33,34], or detrapping of triplet excitons[35]. These models have had considerable success in describing the spin-spin interactions present in these organic systems. Unfortunately, organic systems are disordered, contain an abundance of nuclear spins, and multiple mechanisms could conceivably be involved in many cases. This makes it difficult to develop models from which NZFMR line shapes can be directly linked to defect structures. Inorganic crystalline semiconductors and some inorganic amorphous semiconductors offer substantial advantages in analysis. First, long range order is present in crystalline semiconductors. Second, short range order is often present in amorphous inorganic semiconductors. Third, popular constituent elements silicon and carbon contain few nuclear magnetic moments. Additionally, transport phenomena in inorganic semiconductor devices is generally well understood. We will focus on two models that best align with our understanding of the spin-spin interactions happening in inorganic semiconductor materials and devices in our NZFMR experiments. These models include an adaptation of the e-h pair model[33,34] for the case of the SDR and an adaptation of the two-site model[37] for the case of SDTAT. In order to explain these models, we first need to understand the SLE first proposed by Scully and Lamb.[38]

a. The Stochastic Liouville Equation

Employment of the density-matrix formalism is instrumental to describe the statistical nature of an ensemble of spin states. Consider an ensemble of N spin-pairs in their respective spin states, $\Psi_n$. The dynamics of the ensemble can be described by the density operator

$$\hat{\rho} = \frac{1}{N}\sum_{n=1}^{N} |\Psi_n\rangle\langle\Psi_n|. \tag{3}$$

In a complete orthonormal basis, $\phi_i$, we can then define the density-matrix as

$$\rho_{i,j} = \langle\phi_i|\hat{\rho}|\phi_j\rangle, \tag{4}$$

where elements of $\rho$ describe the probability of finding the spin system in a corresponding basis state. The density-matrix, $\rho$, completely describes the state of the ensemble that includes both electronic and nuclear spins. The time evolution of that ensemble can be described through the Liouville equation (LE)

$$\frac{d\rho}{dt} = -\frac{i}{\hbar}[\mathcal{H}, \rho]. \tag{5}$$

The square brackets represent the commutator operation while $\mathcal{H}$ represents the Hamiltonian of the spin system.

The LE represents the non-dissipative, coherent evolution of the spin ensemble through time, without taking interactions with the environment into account. In order to properly describe both the coherent evolution and the environmentally dissipative, incoherent evolution of the spin ensemble, we turn to the SLE,

$$\frac{d\rho}{dt} = -\frac{i}{\hbar}[\mathcal{H}, \rho] - \frac{k}{2}\{\Lambda, \rho\} + p\Gamma. \tag{6}$$

The first term on the right side of the SLE describes the coherent evolution of spin pairs in the absence of interactions with the environment, exactly as the LE would. The second term introduces a dissipative effect that selectively removes spin pairs from the system at a rate of $k$. The projection operator, $\Lambda$, projects onto the spin subspace from which the projected interactions are removed. The braces in the second term represent the anticommutator. The third term represents a source term that adds spin pairs in random orientations to the system at a rate of $p$ and is typically proportional to an identity matrix, $\Gamma$, in non-magnetic systems.[30]

Equation 6 is able to describe the generation, annihilation, coherent and non-evolution of spin-spin interactions through time.[30] This equation also serves as the basis from which the adaptation of the e-h pair model[33,34] and the adaptation of two-site model[37] are formed that best describe SDR and SDTAT, respectively.

b. Adaptation of the electron-hole pair model

The e-h pair model put forth by Prigodin *et al.*[34] and the trap-induced magnetoresistance model of Harmon and Flatté[33] serve as the basis from which we constructed our model that best aligns with SDR taking place at a deep level defect. In this model, depicted in Figure 1, we consider an electron in the conduction band, a hole in the valence band, and an unpaired electron in a defect such as a dangling bond associated with a nuclear spin and hyperfine field present. The electron in the conduction band drops to an intermediate state. In the absence of a magnetic field, the singlet and triplet spin state between the electron in the intermediate state and the electron in the deep level defect site will be mixed by the hyperfine field of the nucleus. In the case of a singlet pair the electron in the intermediate state will fall to the defect and recombine with the hole in the valence band. Recombination removes these carriers, and therefore spins, from the ensemble and reduces device current. In the case of a triplet pair, the electron in the conduction band has only the option to dissociate from the intermediate state and return to the conduction band. Since the electron is unable to recombine with the hole, device current is not altered in the system. The application of a modest external magnetic field suppresses the mixing caused by the hyperfine fields.[30] As the modest external field is reduced, the recombination rate increases.

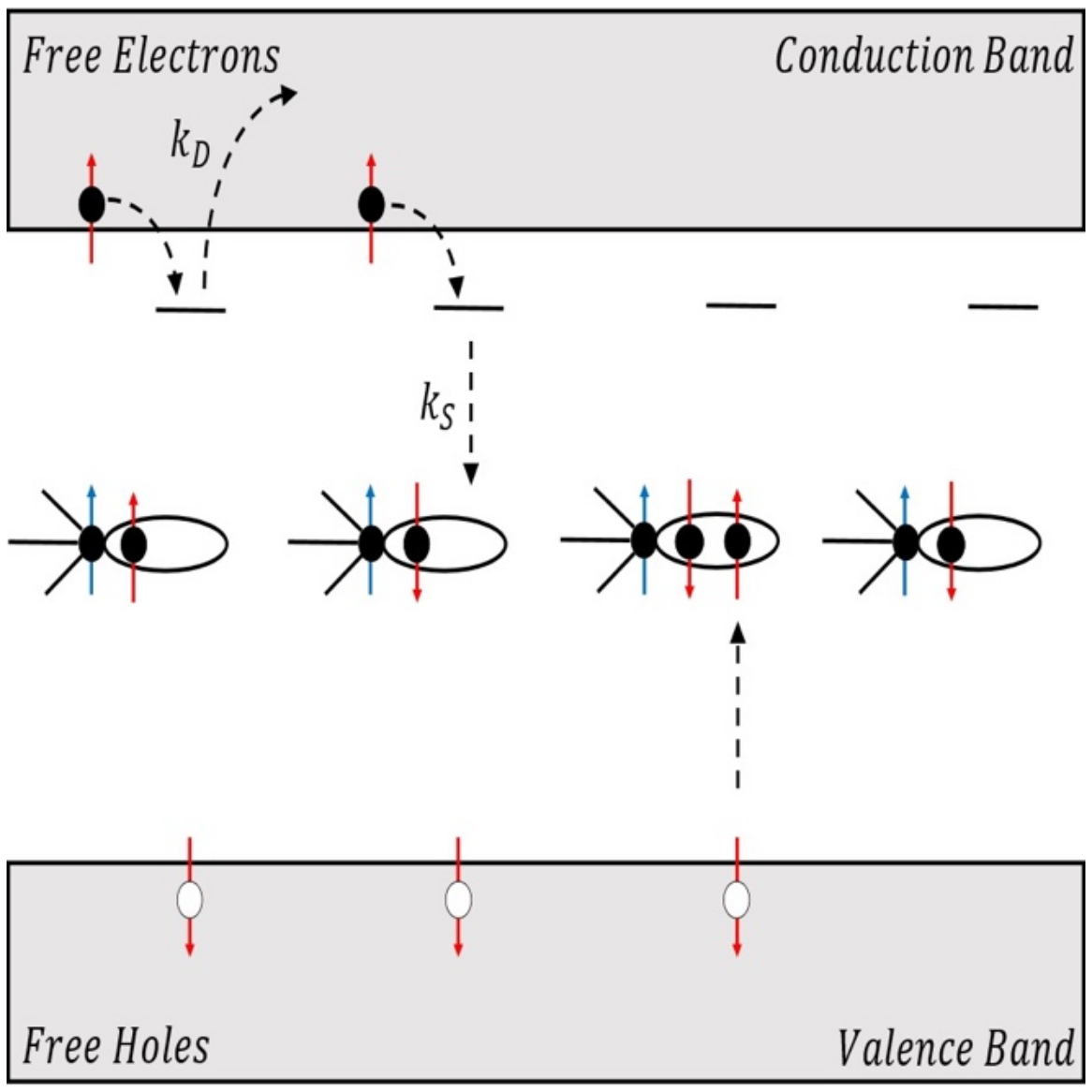

Figure 1: Spin dependent recombination model.

The time evolution of a density-matrix that can best describe this spin system can be represented by an adaptation of the SLE proposed by Hansen and Pedersen[39]:

$$\frac{d\rho}{dt} = -\frac{i}{\hbar}[\mathcal{H},\rho] - \frac{1}{2}(k_S + k_D)\{\Lambda_S,\rho\} - \frac{k_D}{2}\{\Lambda_T,\rho\} + \frac{1}{8}p\Gamma. \quad (7)$$

The first term on the right side describes the coherent evolution of the spin pairs in the absence of the interaction with the environment - the mixing of the states caused by the hyperfine fields. The second term is a dissipative effect that selectively removes singlets (e-h pair) from the system at a rate of $k_S$ and the third term is also a dissipative term that represents the selective removal of triplets that have dissociated to the conduction band at a rate of $k_D$. $\Lambda_S$ and $\Lambda_T$ denote the singlet and triplet projection operators, respectively. The fourth term represents a source of carriers to the system that are generated a rate of $p$ while $\Gamma$ represents an identity matrix. Since carriers are being generated and annihilated at constant rates, we are only concerned with the steady state form of the SLE ($dp/dt = 0$). The resulting form of the SLE is then

$$0 = -\frac{i}{\hbar}[\mathcal{H}, \rho] - \frac{1}{2}(k_S + k_D)\{\Lambda_S, \rho\} - \frac{k_D}{2}\{\Lambda_T, \rho\} + \frac{1}{8}p\Gamma. \quad (8)$$

The factor of $\frac{1}{8}$ in the last term exists because the dimension of $\Gamma$ is 8 such that the total generation rate of the spin pairs is $\frac{1}{8}\mathrm{pTr}(\Gamma) = p$ which is independent of the spin dimension.

The Hamiltonian for a single deep level recombination site within the bandgap can be represented by

$$\mathcal{H} = g\mu_B(\boldsymbol{S_1} + \boldsymbol{S_2}) \cdot \boldsymbol{B} + a(\boldsymbol{S_2} \cdot \boldsymbol{I}), \quad (9)$$

where the first term represents the Zeeman splitting of the electron spins, $\boldsymbol{S_1}$ (electron in the conduction band) and $\boldsymbol{S_2}$ (electron in the defect site), in the magnetic field while the second term represents an isotropic electron-nuclear hyperfine interaction, of strength $a$, between the electron, $\boldsymbol{S_2}$, and the nucleus, $\boldsymbol{I}$. In this model, we assume that the electron in the intermediate state is not interacting with any nuclear spin or hyperfine field.

c. Adaptation of the two-site Model

The two-site model for organic semiconductors put forth by Wagemans *et al.* that considers only two characteristic sites best aligns with our current understanding of SDTAT in dielectrics and insulators.[37] SDTAT is enabled through variable range hopping and exploits the conservation of spin angular moment from trap-to-trap tunneling events.[9] Despite the existence of many sites that a carrier may occupy while hopping through the disordered system, the resistance is dominated by one bottleneck pair for which the two-side model appropriately describes.[32] Figure 2, depicts the spin-dependence of the trap-assisted tunneling event. In order for the electron to successfully tunnel or hop to the other site, the two electron spins must be in a singlet configuration. If they are in a triplet configuration, the hopping event is forbidden. As the resonance condition is reached and SDTAT is detected through EDMR, the transition from triplet to singlet configurations is forced, allowing for previously forbidden tunneling events to take place, and thus a change in leakage current is measured. SDTAT detected though NZFMR does not involve a resonance induced transition of the spin state but rather a mixing of states.[10,22,40]

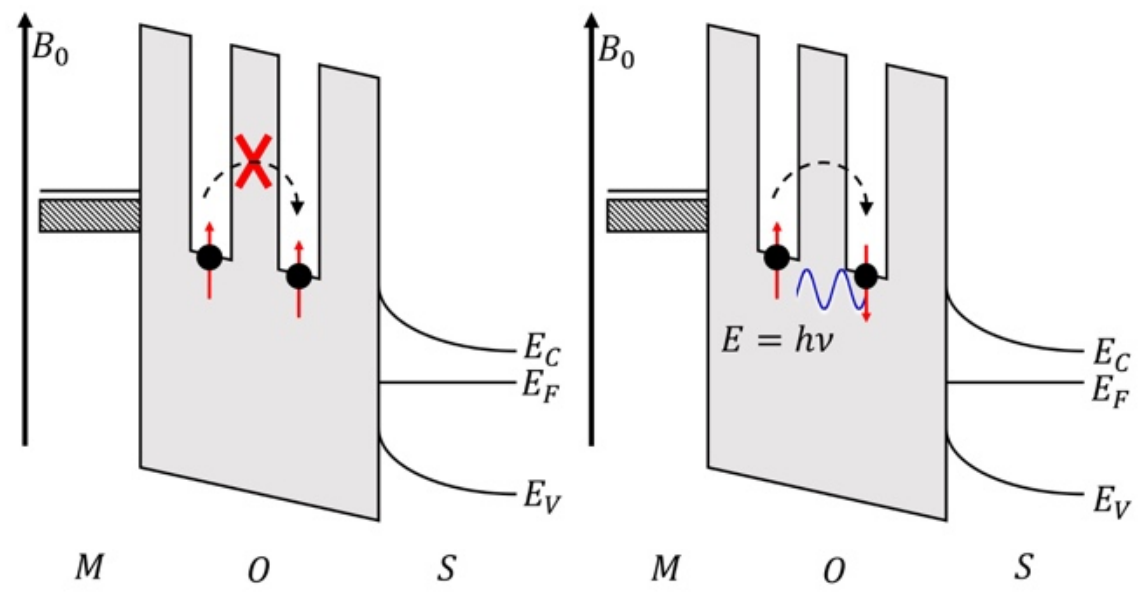

Figure 2: Depiction of SDTAT. Tunneling between traps is forbidden due to the triplet configuration (left). Tunneling is allowed due to the triplet configuration (right).

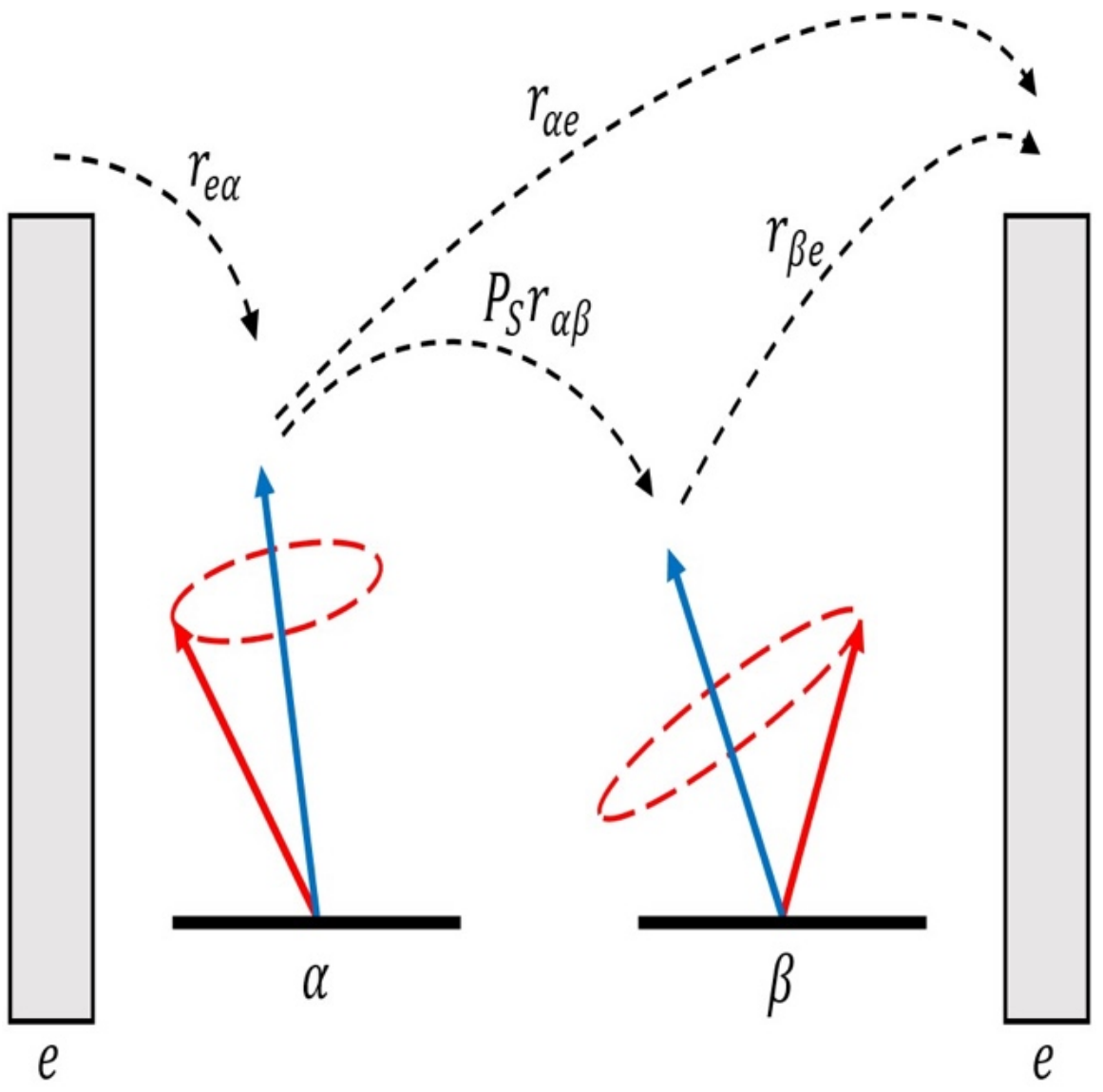

Figure 3: Depiction of the two-site model proposed by Wegemans *et al.*[37] Image adapted from Schellekens *et al.*[30]

The adapted two-site model, as depicted in Figure 3, consists of two nuclear sites, $\alpha$ and $\beta$. Both sites have their own respective nuclear spins (both spins assumed $I = ½$), and hyperfine fields and thus coupling constants, $a_\alpha$ and $a_\beta$, when singly occupied with an electron. If site $\alpha$ is unoccupied, an electron can hop from the environment, $e$, to site $\alpha$ at a rate of $r_{ea}$. Upon the occupation of site $\alpha$, the electron will precess about the hyperfine field of the present nucleus. At this point, the electron can either hop directly to the environment at a rate of $r_{ae}$ or hop to site $\beta$ at a rate of $P_s r_{ab}$, where $P_s$ is the probability of the electron at site $\alpha$ and the electron at site $\beta$ being in a singlet configuration. After occupation of site $\beta$, the electron can dissociate back into the environment at a rate of $r_{be}$. In this model, the current flow is defined as the total flow of electrons from the environment to site $\alpha$. In order to greatly simplify the arithmetic, we assume that the current limiting rate is not the electron hopping from site $\beta$ into the environment, therefore we can assume $r_{be} = \infty$.[30]

The density-matrix representation of this ensemble can be represented by an adapted SLE as originally proposed by Wagemans *et al.*[37],

$$\frac{d\rho}{dt} = -\frac{i}{\hbar}[\mathcal{H},\rho] - \frac{1}{2}(r_{ab} + r_{ae})\{\Lambda_S,\rho\} - \frac{1}{2}r_{ae}\{\Lambda_T,\rho\} + \frac{1}{16}r_{ea}(1 - Tr(\rho))\Gamma, \qquad (10)$$

where $\Lambda_S$ and $\Lambda_T$ are the singlet and triplet project operators, respectively, and $\Gamma$ is an identity matrix that corresponds to a generation of spins at a random orientation. The current through the system is the current through site $\alpha$ and is defined as: $I = r_{ea}\big(1 - Tr(\rho)\big)$. It should be noted that since we are interested in steady state conditions, the current though the system is constant and therefore the current, $I$, is treated as a constant scalar value. Assuming steady state conditions ($dp/dt = 0$) and that traps remain mostly unoccupied ($1 \gg Tr(\rho)$), equation (10) becomes

$$0 = -\frac{i}{\hbar}[\mathcal{H}, \rho] - \frac{1}{2}(r_{ab} + r_{ae})\{\Lambda_S, \rho\} - \frac{1}{2}r_{ae}\{\Lambda_T, \rho\} + \frac{1}{16}r_{ea}\Gamma. \quad (11)$$

Since there are now four spins included, the dimension of $\Gamma$ is $16$ which leads to the factor of $\frac{1}{16}$ in the last term in the same way we had for spin dependent recombination involving three spins.

The Hamiltonian of this system can be represented by

$$\mathcal{H} = g\mu_B(\boldsymbol{S_1} + \boldsymbol{S_2}) \cdot \boldsymbol{B} + a_\alpha(\boldsymbol{S_1} \cdot \boldsymbol{I_\alpha}) + a_\beta(\boldsymbol{S_2} \cdot \boldsymbol{I_\beta}), \quad (12)$$

where the first term represents the Zeeman splitting of the electron spin, $\boldsymbol{S_1}$ and $\boldsymbol{S_2}$, while the second and third term represents an isotropic electron-nuclear hyperfine interaction, of strength $a$, between the nucleus and the electron occupying the site.

d. Solution of the SLE

In both models, we solve for the steady state ($dp/dt = 0$) where the equations can be algebraically manipulated to take the form:

$$\mathcal{L}\rho + \rho\mathcal{L}^\dagger = g, \quad (13)$$

where

$$\mathcal{L} = \frac{i}{\hbar}\mathcal{H} + \frac{(c_1+c_2)}{2}\Lambda_S + \frac{c_2}{2}\Lambda_T. \quad (14)$$

Here, $c_1$ and $c_2$ represent the respective rate constants that depend upon the form of the SLE being solved and $g$ represents the generation rate at which spins are being introduced into the system. Equation 13 is in the form of the time-continuous Lyapunov equation. The solution of Equation 13 and the density-matrix at steady state can be represented as

$$\rho = \int_0^\infty e^{-i\mathcal{L}t} g e^{i\mathcal{L}^\dagger t} dt. \quad (15)$$

In both models, the measurable-observable change in current is due to singlet-related events, which makes us concerned with the singlet projection of the density-matrix. The measurable SDR current is therefore proportional to the product of the probability of the formation of a singlet and the trace of the singlet population of the density-matrix,

$$I_{SDR} \propto Z\left(\frac{k_S}{p}\right)Tr(\Lambda_S\rho). \tag{16}$$

Here the constant $Z$ describes the relative contribution due a signal in terms of an amplitude scaling factor. The measurable SDTAT current is proportional to product of the rate at which the sites are filled and the probability of occupation of the sites,

$$I_{SDTAT} \propto r_{ae}\big(1 - Tr(\rho)\big). \tag{17}$$

III.) Least Squares Fitting of the SLE

When solving for the density-matrix in Equation 15, $\rho$ is an inherent function of the rate constants, $c_1$ and $c_2$, the generation rate, $g$, and the Hamiltonian, $\mathcal{H}$. The Hamiltonian is a function of the externally applied magnetic field and hyperfine coupling constant(s). We know the externally applied magnetic field to the system at each point in our measurement, so the Hamiltonian simply becomes a function of the hyperfine coupling constant ($a$ in the case of SDR or $a_\alpha$ and $a_\beta$ in the case of SDTAT). If we consider the simplest case of a single recombination center with a hyperfine coupling constant, $a$, in the case of SDR, or if we consider two sites with the same hyperfine coupling constant, $a = a_\alpha = a_\beta$, in the case of SDTAT, then the density-matrix becomes a function of hyperfine coupling constant, $a$, rates $c_1$ and $c_2$, and generation rate, $g$, so

$$\rho = \rho(a, c_1, c_2, g). \tag{18}$$

Since the measurable-observable current, $I$, through the system is proportional to the density-matrix, we find that the measurable-observable current is a function of those same constants,

$$I = I(a, c_1, c_2, g). \tag{19}$$

Given a measured NZFMR spectrum as function of magnetic field, we can then fit a simulated spectrum to the given data set using the appropriate form of the SLE via a non-linear weighted least squares method. The residual error, $r^2$, of the least squares method can be defined as

$$r^2(a, c_1, c_2, g) = \sum_i \left( w_i \left( I_{M,i} - I_{S,i}(a, c_1, c_2, g) \right)^2 \right), \tag{20}$$

where $w_i$ is a weighting factor. Here the residual square error between the measured spectrum, $I_M$, and the simulated spectrum, $I_S$, is a function of the hyperfine coupling constant $a$, rates $c_1$ and $c_2$, and generation, $g$. When the residual error is minimized at some set of values, that is when

$$\nabla\left(r^2(a, c_1, c_2, g)\right) = 0, \tag{21}$$

we can gain the information about the defects present as a result of the value corresponding to hyperfine coupling constant(s) at the minimum.

In our fitting algorithm, we defined the weights, $w_i$, on the curvature of the spectra (second derivative of the absorption spectra) such that

$$w_i = \left| \left( \frac{dI_M}{dB_0} \right)_i \right|. \tag{22}$$

The second derivative of the NZFMR spectrum highlights areas where important features generally appear. Weighting from the second derivative places higher residual error, or effectively higher importance, on these features for fitting. This forces the minimization routine to fit these sections of the experimental data with greater accuracy instead of neglecting them.

Equation 20 is minimized through an interior point minimization routine outlined by Byrd *et al.*[41] This type of algorithm is also built into the function fmincon within the global optimization toolbox in MATLAB. The minimization routine is also passed a window of constraints for the values the hyperfine interactions. We know that these values generally fall in between 0 and 100 mT. Anything larger than 100 mT for the material systems we choose to investigate is physically unrealistic. The rates $c_1$ and $c_2$, and generation rate, $g$, were left unconstrained.

## IV.) Experimental Details

Our fitting algorithm was applied to the NZFMR spectrum of two distinct material systems: $Si/SiO_2$ MOSFETs and a-Si:H MIS capacitors. Experimental details and biasing schemes varied for each material system.

The NZFMR measurement of the $Si/SiO_2$ device was conducted via the DCIV (gated-diode) biasing scheme where the source and drain are shorted together and forward biased and the gate is biased so that the channel resides in depletion.[42] When the gate bias is swept through depletion, a peak in substrate current is measured. This peak in substrate current is due to recombination at interfacial defects and is observed when there are roughly equal populations of electrons and holes are present at the interface. The DCIV NZFMR response is measured via the substrate current, with the gate bias held at the peak recombination current. This biasing scheme has been used in the past to explore the $Si/SiO_2$ interface via DCIV EDMR.[43] In our case, the gate bias was 0.3V, and the source and drain bias was -0.33V. This device was irradiated with a $^{60}$Co source to a dose of 1Mrad(Si) in order to generate interface states. The NZFMR DCIV measurement reported was made after irradiation. No detectable NZFMR DCIV signal was present before irradiation. The $Si/SiO_2$ MOSFET used had an oxide thickness of 7.5 nm and a gate area of 41,000 $\mu m^2$.

The NZFMR measurements on a-Si:H MIS capacitor was made via SDTAT. The a-Si:H was biased with -2 V and the SDTAT NZFMR response was measured via the leakage current across the insulator region.[44] The MIS stack consisted of Ti/10 nm of a-Si:H/p-Si.

## V.) Results and Discussion

### a. Spin Dependent Recombination (SDR)

The top of Figure 4 depicts a gated diode DCIV NZFMR measurement of a gamma irradiated Si/$SiO_2$ MOSFET and its fitted spectrum. Based on EDMR measurements, the MOS interface defects involved are silicon dangling bonds, $P_b$ centers.[45–48] We would also expect a fair amount of hydrogen in the interface region due to the irradiation damage[49], so we would also expect the possibility of an interaction with nearby hydrogen. We constructed our fitting algorithm to consider two separate recombination paths each with their own isotropic hyperfine coupling constant acting as an adjustable parameter with which to fit the experimental data. The first recombination path describes an electron falling from the intermediate state to a deep level $P_b$ center that is interacting with distant $^{29}$Si while the second path describes an electron falling from the intermediate state to a deep level $P_b$ center that is interacting with nearby hydrogen. The recombination and dissociation rates were tethered together since we assume that these rates should be the same at all $P_b$ centers regardless of the interaction taking place with the $P_b$ center. The scaling factors that relate to the relative contributions were left as independent adjustable parameters. This brought the total number of adjustable parameters to six. The algorithm then fit these six parameters to the experimental data. Table 1 contains the isotropic hyperfine coupling constants, $a$, singlet recombination rates, $k_S$, pair dissociation rates, $k_D$, and relative contribution scaling factor, $Z$, the fitting algorithm produced for the two recombination paths.

Our fitting algorithm generated the hyperfine coupling constants to be 1.4 mT and 0.2 mT. An interaction distant $^{29}Si$ interaction with a $P_b$ center has an isotropic hyperfine coupling constant that ranges between 1.4-1.6 mT.[50–52] We find the 1.4 mT electron-nuclear hyperfine constant generated as result of the fit to be consistent with interactions between $P_b$ centers with distant $^{29}Si$ while we attribute the 0.2 mT to an interaction with nearby hydrogen.

We are also confident of the parameters listed in Table 1 to be within roughly ±20%. Figure 5 displays the percent difference from the minimum square error between the generated spectrum and the experimental spectrum as a range in colors within the parameter space. The axes of plot represent the parameter space and range from 80% to 120% of the values of best fit for the singlet recombination rates, $k_S$, pair dissociation rates, $k_D$, and isotropic hyperfine coupling constants, $a$, for the x, y, and z axes, respectively. Two plots were generated such that one hyperfine parameter of the two was fixed at its value of best fit while the second hyperfine parameters varied. The centers of the figures, (100%,100%,100%), are the values of best fit as presented in Table 1 and display the minimum square error. Points further from the center display increased minimum square error. The outer most corners of the plot on the top and bottom planes represent the extremes and of the parameter ranges and display differences from minimum square errors of 100% of the minimum. We attribute the larger range in the difference from the minimum square error to the modest signal-to-noise ratio of the experimental spectrum. The algorithm's primary task is to reduce the residual error between the experimental data and its fit. Lower SNR inherently forces more residual error into the fit, decreasing the confidence of the values of best fit.

From the ratio of the relative contribution scaling factors, we found the relative abundances of the distant $^{29}Si$ and nearby hydrogen to be about 7% and 93% respectively. This is consistent with what we would expect at the interface; while the amount of hydrogen is not precisely known, it is reasonable to assume it to be much greater than the number of $^{29}$Si sites.

In a recent publication[53], Si/$SiO_2$ NZFMR and EDMR were examined within a mean field model where hyperfine interactions were assumed to be due to a Gaussian distribution of classical nuclear spin vectors at each of the electron sites (*i.e.* shallow and deep states). The work found a typical hyperfine field magnitude at the deep level defect to be 0.5 mT which falls within the range of values determined here. A complete comparison between the two approaches will require the model presented herein to be extended to include EDMR.

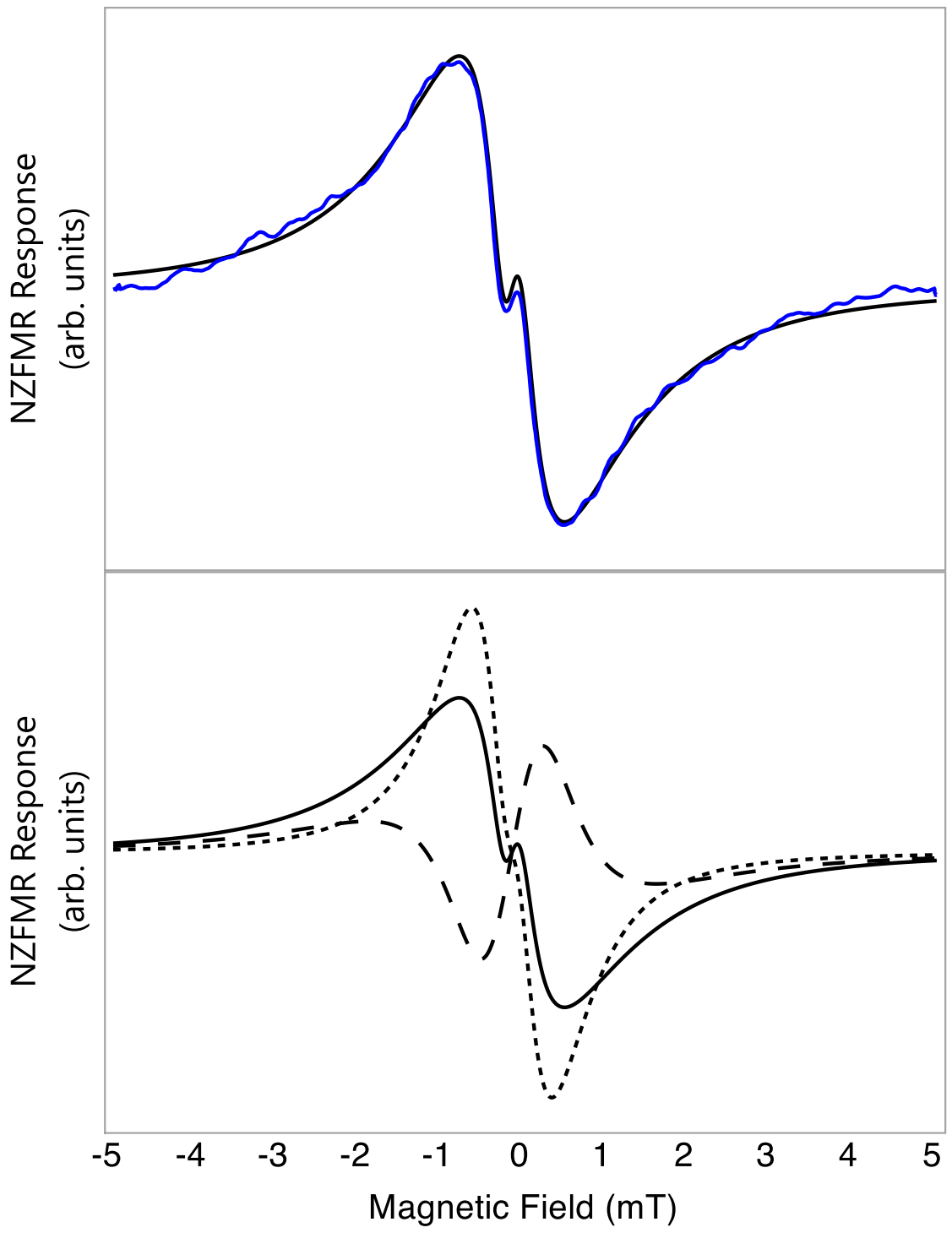

Figure 4: Si/$SiO_2$ MOSFET DCIV NZFMR experimental spectrum (blue). Fitted NZFMR spectrum (solid black) that is the superposition of the spectra due to a $P_b$ center electron-nuclear hyperfine interactions with distant $^{29}$Si (long dash) and nearby Hydrogen (short dash).

Table 1: Fitted parameters of the SLE to the DCIV NZFMR Measurement

| $P_b$ Interactions with: | $a$ (mT) | $k_S$ (GHz) | $k_D$ (GHz) | $Z$ (arb. units) |
|---|---|---|---|---|
| Distant $^{29}Si$ | 1.401 | 0.174 | 0.037 | 24.333 |
| Nearby Hydrogen | 0.206 | 0.174 | 0.037 | 1.704 |

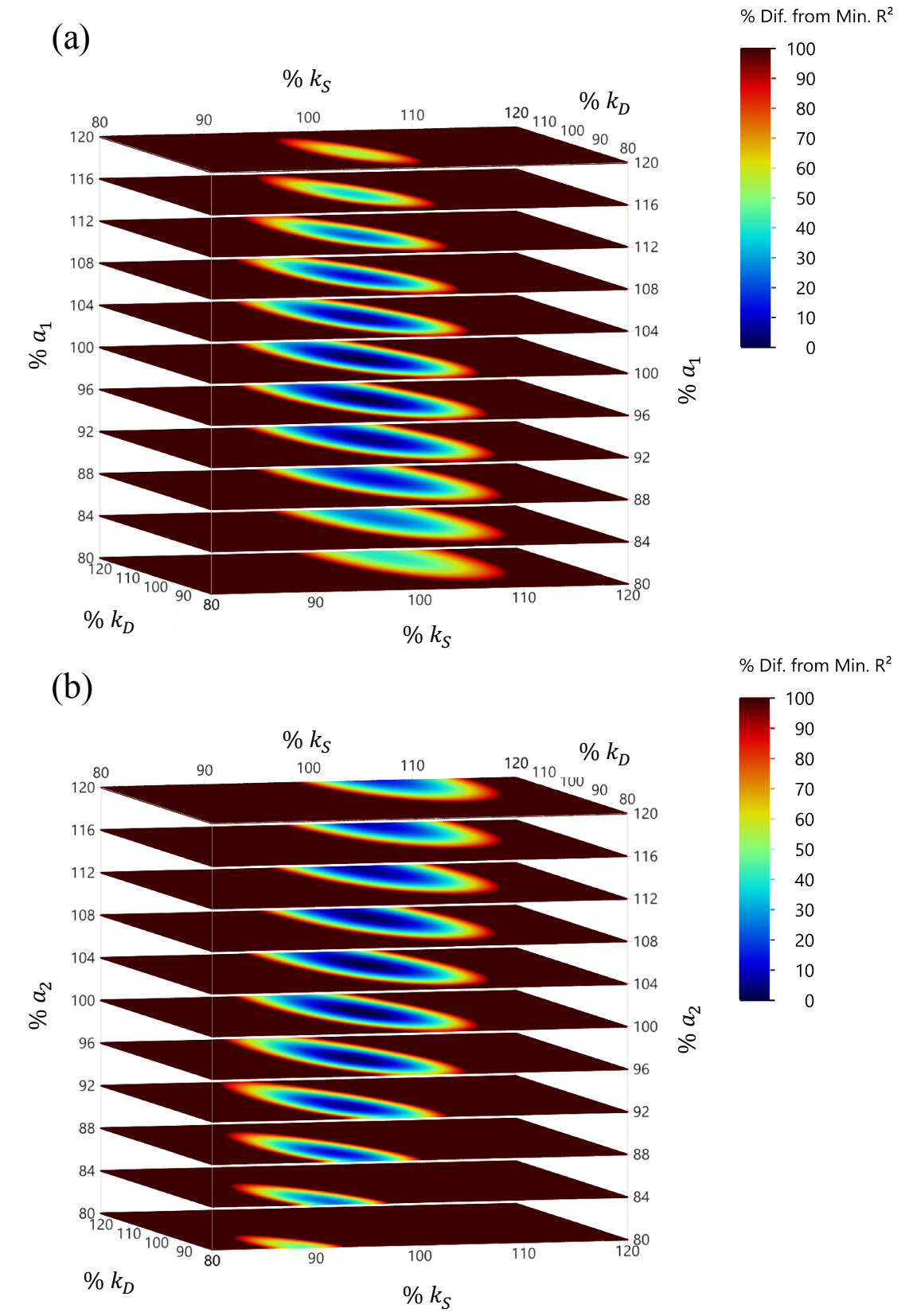

Figure 5: Percent difference from the minimum square error displayed as a color as a function of the parameter space for Si/$SiO_2$ MOSFET DCIV NZFMR measurement. a.) $a_1$ was varied while $a_2$ was held constant the value of best fit. a.) $a_2$ was varied while $a_1$ was held constant the value of best fit.

b. Spin Dependent Trap Assisted Tunneling (SDTAT)

The top of Figure 6 shows an NZFMR SDTAT measurement in a-Si:H along with a fitted spectrum. Based on previous high-field EDMR measurements and as well as EPR and EDMR literature, the dominating defect is the silicon dangling bond. As reported in the literature, a distant $^{29}Si$ interaction with a silicon dangling bond in a-Si:H has an isotropic hyperfine coupling constant of about 1.5 mT with an integrated intensity corresponding to a relative ratio of abundance of about 14%.[54] These particular a-Si:H samples also had a very high concentration of hydrogen of about 20-30%, so we expect the dangling bond to have some hyperfine interaction with nearby hydrogen. The coupling constant for hydrogen ranges between 0.35 mT and 1.0 mT (potentially higher) depending upon average distance between hydrogen and the dangling bond.[55,56] Since interactions with nearby hydrogen are much more likely than interactions with distant $^{29}Si$, the fitting algorithm was constructed to consider two separate hopping paths. The first path (labeled Path 1 in Table 2) describes an electron hopping from a dangling bond interacting with nearby hydrogen to another dangling bond also interacting nearby hydrogen. We would expect that the isotropic hyperfine coupling constant at both sites interacting with hydrogen would be same, that is we expect that $a_\alpha = a_\beta$. The second path (labeled Path 2 in Table 2) describes an electron hopping from a dangling bond interacting with nearby hydrogen to another dangling bond experiencing an interaction with distant $^{29}Si$. These two sites in this path are experiencing two different hyperfine interactions, so we cannot assume their isotropic coupling constants would the same, that is $a_\alpha \neq a_\beta$. Since all of the interactions with nearby hydrogen must have the same isotropic hyperfine coupling constants, all three parameters were tethered together as one adjustable parameter. The isotropic hyperfine coupling constant for distant $^{29}Si$ was left as another adjustable parameter. We also assume the probability of hopping rates must be the same between paths. That is $r_{ae}$ and $r_{ab}$ must be the same for both paths, so $r_{ae}$ values were tethered together as one adjustable parameter while $r_{ab}$ values were tethered together as another adjustable parameter. The rate at which the two sites are filled, $r_{ea}$, were left as separate independent adjustable parameters. This brought the total number of adjustable parameters to six. The algorithm then fit these six parameters to the experimental data. Table 2 contains the isotropic hyperfine coupling constants, $a_\alpha$ and $a_\beta$, hopping rates, $r_{ae}$ and $r_{ab}$, and the rate at which the two sites were filled, $r_{ea}$, that the fitting algorithm produced for the two hopping paths. Our fitting algorithm generated the hyperfine coupling constants for the two interactions present in the hopping paths to be about 1.7 mT and 0.3 mT. We find these isotropic electron-nuclear hyperfine interactions to be consistent with dangling bond interactions with distant $^{29}$Si and nearby hydrogen.

Given that the electron starts at a site interacting with nearby hydrogen, the probability that the electron will hop to a site interacting with distant $^{29}$Si is about 15%. The probability that the electron will hop to another site interacting nearby hydrogen is 20-30%. Therefore the probability of hopping to a site experiencing an interaction with nearby hydrogen is about 1.3-2 times more likely than hopping to a site experience an interaction with distant $^{29}$Si. The ratio of rates at which the site are filled from Table 2, $r_{ea}$, indicate that the hop to a site interacting with nearby hydrogen is 2.6 times more likely than a hop to a site experience an interaction with distant $^{29}$Si. We find agreement between the ratio of the rates at which the sites are filled to be consistent with the two most likely hopping paths.

We are confident of the parameters listed in Table 2 to be within at least ±10%. Figure 7 displays the percent difference from the minimum square error of a generated spectrum and the experimental data as a range in colors within the parameter space. The axes of plot represent the parameter space and range from -10% to +10% of hopping rates, $r_{ae}$ and $r_{ab}$, isotropic hyperfine coupling constants, $a$, for the x, y, and z axes, respectively. Two plots were generated such that one hyperfine parameter of the two was fixed at its value of best fit while the second hyperfine parameters varied The centers of the figures, (100%,100%,100%), are the values of best fit as presented in Table 1 and display the minimum square error. Points further from the center display increased minimum square error. The outer most corners of the plot on the top and bottom planes represent the extremes and of the parameter ranges and display differences from minimum square errors off 100% of the minimum. We attribute the smaller range in the difference from the minimum square error to the higher signal-to-noise ratio (SNR) of the experimental spectrum. Since SNR was significantly higher in for this experimental data, the confidence of the fit was higher as well.

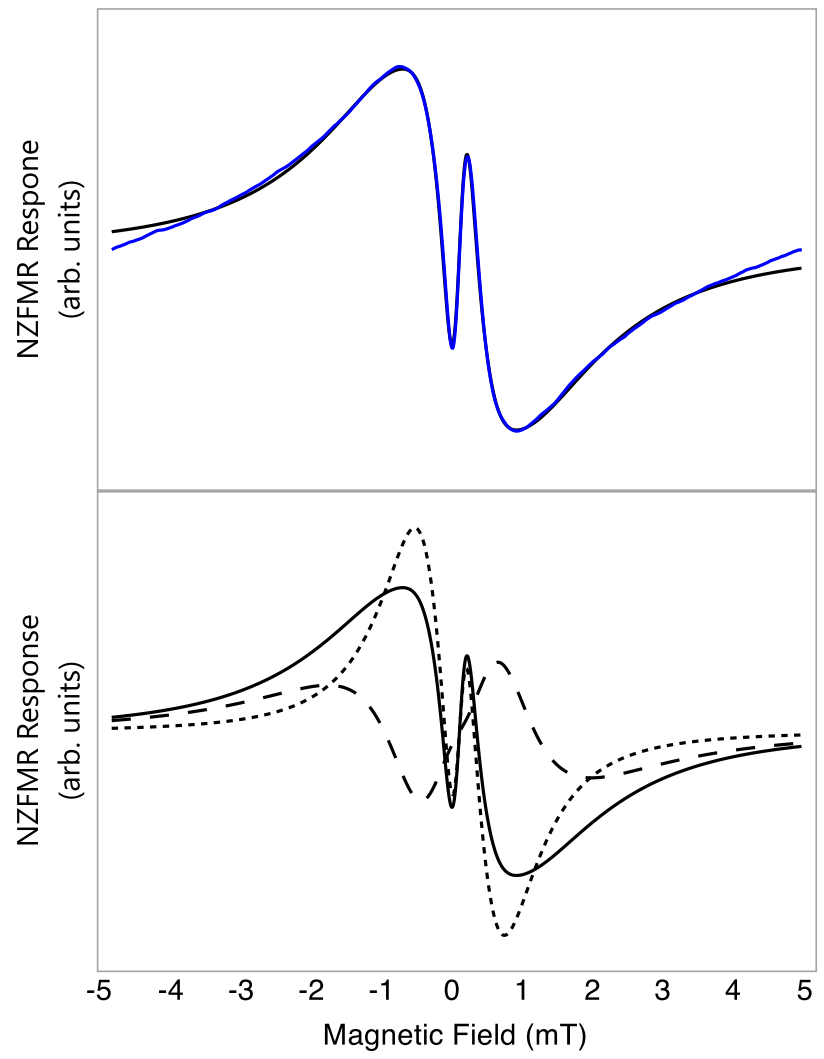

Figure 6: a-Si:H SDTAT NZFMR experimental spectrum (blue). Fitted NZFMR spectrum (solid black) that is the superposition of the spectra due to hopping from site A to site A (short dash) and hopping from site A to Site B (long dash).

Table 2: Fitted parameters of the SLE to the BAE NZFMR Measurement

| Hopping Paths | $a_\alpha$ (mT) | $a_\beta$ (mT) | $r_{ab}$ (GHz) | $r_{ae}$ (GHz) | $r_{ea}$ (GHz) |
|---|---|---|---|---|---|
| Path 1 | 0.332 | 0.332 | 0.128 | 0.048 | 616.488 |
| Path 2 | 0.332 | 1.721 | 0.128 | 0.048 | 238.244 |

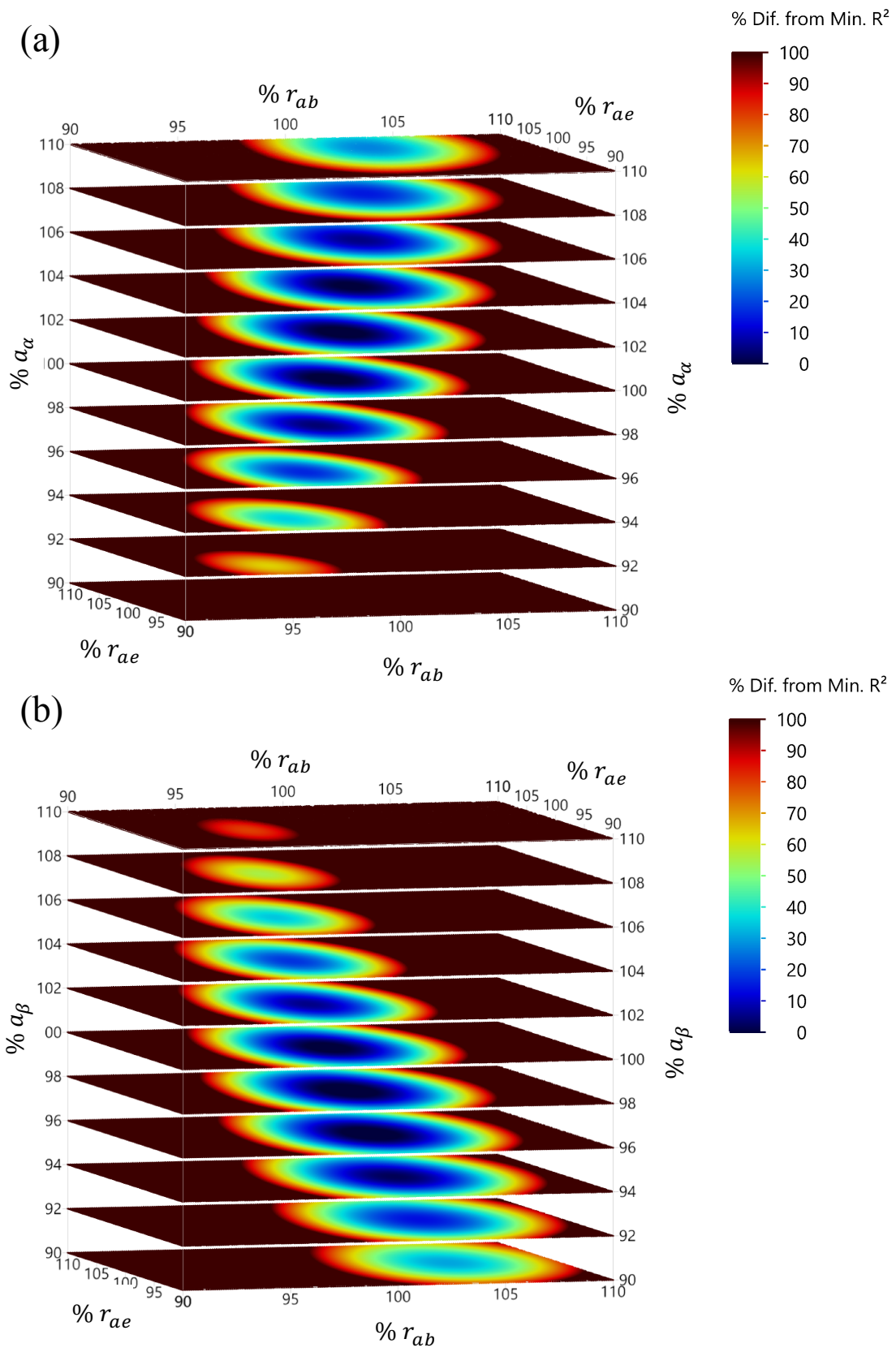

Figure 7: Percent difference from the minimum square error displayed as a color as a function of the parameter space for a-Si:H SDTAT NZFMR measurement. a.) $a_\alpha$ was varied while $a_\beta$ was held constant the value of best fit. a.) $a_\beta$ was varied while $a_\alpha$ was held constant the value of best fit.

## VI.) Conclusions

In this work we show that the NZFMR response contains at least some of the analytical power that has been demonstrated in EPR and EDMR measurements when determining the chemical and physical nature of point defects in semiconductor devices and insulator thin films. NZFMR measurements are straightforward to conduct, and they do not require complex microwave circuity. The NZFMR measurement is also expected to be possible on devices deep within integrated circuits, where EPR and EDMR would not be possible. We discuss the use of two models based on the SLE used to describe the NZFMR response in semiconductors and insulators. With relatively small computational effort, we fit the experimental NZFMR response from two distinct materials systems by utilizing knowledge of the present spin systems, the SLE, and a least squares fitting approach. Spin-spin interaction strengths, such as isotropic electron-nuclear hyperfine coupling constants, are extracted from each system, and are in good agreement with prior knowledge of the defects present in the systems. Our work indicates that the NZFMR response and fitting of the NZFMR spectrum via the SLE could be a relatively simple yet powerful addition to the family of spin-based techniques used to explore the chemical and physical nature of point defects in semiconductor devices and insulator thin films. We note that performing a simultaneous fit of the results obtained with NZFMR with those from EDMR on the same device may provide still more insight into the effectiveness of NZFMR as an independent tool.

Acknowledgments

This work was supported by the Defense Threat Reduction Agency (DTRA) under award number HDTRA1-18-0012. The content of the information does not necessarily reflect the position or the policy of the federal government, and no official endorsement should be inferred.

## Data Availability

The data that support the findings of this study are available from the corresponding author upon reasonable request.